\documentclass[twocolumn]{aastex631}
\usepackage{fix-cm}

\usepackage{amsmath}
\usepackage{soul}
\usepackage{hyperref}

\usepackage{CJK}

\shorttitle{Magnetically Arrested Circumbinary Accretion Flows}
\shortauthors{Most \& Wang}
\graphicspath{{./}{figures/}}

\begin{document}
\begin{CJK*}{UTF8}{gbsn}

\title{Magnetically Arrested Circumbinary Accretion Flows}

\author[0000-0002-0491-1210]{Elias R. Most}
\email{emost@caltech.edu}
\affiliation{TAPIR, Mailcode 350-17, California Institute of Technology, Pasadena, CA 91125, USA}
\affiliation{Walter Burke Institute for Theoretical Physics, California Institute of Technology, Pasadena, CA 91125, USA}
\author[0000-0001-7167-6110]{Hai-Yang Wang(王海洋)}
\email{haiyangw@caltech.edu}
\affiliation{TAPIR, Mailcode 350-17, California Institute of Technology, Pasadena, CA 91125, USA}

\begin{abstract}
    Binary systems with comparable masses and a surrounding accretion disk can accrete gas through spiral accretion streams penetrating the central cavity formed by tidal interactions.
    Using three-dimensional Newtonian magnetohydrodynamics simulations, we investigate the possibility of a magnetically arrested accretion flow through the cavity.
    Rather than solely continuously feeding the binary through spiral accretion streams, the accretion is regulated by the strong magnetic field inside the cavity. 
    Transport of mass and angular momentum onto the binary then proceeds largely periodically in magnetic flux eruption episodes.
    The ejected flux tubes carry angular momentum outwards and away from the binary,
    inject hot plasma into the disk and can launch flares.
    This likely intermittent scenario could have potential implications for the emission signatures of supermassive black hole binaries, and shed light onto the role magnetic fields play in the binary's orbital evolution.
\end{abstract}

\keywords{}

\section{Introduction}

Binary systems are a prime focus of study in astronomy and astrophysics,
including protoplanetary disks \citep{2011ARA&A..49...67W}, binary stars \citep{2008ApJ...681..375K,2024ApJ...964..133M}, and
supermassive black hole binaries \citep{Begelman1980,2001ApJ...563...34M,2005ApJ...622L..93M,2007MNRAS.379..956D,2007Sci...316.1874M}.
Accretion onto the binary has the potential to power electromagnetic
transients and alter the dynamics of the orbit (see, e.g., \citet{Lai2023} for a recent review).
In the case of supermassive black hole binaries, accretion onto the system
may help explain hardening the binary beyond the parsec separation scale
\citep{Begelman1980}. As such, understanding gas accretion onto a supermassive black
hole binary is of crucial importance for understanding the merging black
hole population seen with the Pulsar Timing Array (PTA) \citep{Nanograv2023a,Nanograv2023b} and the Laser Interferometer Space Antenna (LISA)
\citep{lisa2023}.

For binaries surrounded by an accretion disk, the time-varying
gravitational potential of the binary exerts a torque on the surrounding circumbinary disk, balancing viscous torques and opens up a cavity
at 2-3 times the binary separation (near the location of the inner Lindblad resonance \citep{Artymowicz:1994bw,Artymowicz1996,Goldreich:1979zz}).
One of the main features of circumbinary accretion is that accretion proceeds mainly along two spiral arms penetrating the low-density cavity.
A main driver behind circumbinary disk research is to understand the amount of
angular momentum transferred onto the binary, and in particular whether the binary hardens or
widens due to gas accretion (see, e.g., \citealt{Munoz2019,Lai2023} and references therein), as well as to unravel potential electromagnetic
counterparts (e.g., \citealt{Bogdanovic2022, Charisi2022}).

Over the past decade, several numerical studies have clarified many of the hydrodynamical
aspects of circumbinary accretion, including the dependence on binary properties (mass ratio \citep{Duffell2020,Derdzinski:2020wlw,Siwek2023b,Siwek2023a}, 
orbital eccentricity \citep{DOrazio2021,Siwek2023b,Siwek2023a}) and the properties of the disk (aspect ratio \citep{Tiede2020,Dittmann2023b}, viscosity \citep{Dittmann2023,Dittmann2023b}, cooling \citep{Sudarshan2022,Wang2022,Wang2023,Pieren2023}, self-gravity of the gas \citep{Franchini2021,Bourne2023}
). 
The stage when gravitational wave emission is non-negligible has also been studied by considering relativistic precession \citep{DeLaurentiis2024} and decoupling on the final orbits \citep{Tsokaros:2022hjk,Dittmann2023}. 

Due to the long integration times required to reach a quasi-steady state of angular momentum transfer between the binary and the disk \citep{Miranda2017,Munoz2019},
most simulations have been carried out in two-dimensions focusing on the orbital plane.
Few studies have investigated the hydrodynamial torque evolution in
full three dimensions \citep{Moody2019,Bourne2023}.\\
On black hole horizon scales several works
have also investigated the dynamics of the black hole-gas interaction in 
numerical or post-Newtonian relativity \citep{Noble:2012xz,Avara:2023ztw,Farris:2012ux,Gold:2014dta}, including jet launching and mini-disks \citep{Bowen:2016mci,Bowen:2017oot,Armengol:2021shd,Combi:2021dks,Gutierrez:2021png}, which form around the black holes within
their respective Hill spheres \citep{Paschalidis:2021ntt,Bright:2022hnl}.
On arbitrary -- and in the supermassive black context (sub-)parsec -- scales
few simulations have been carried focusing mainly on the outer disk, and
its magnetic torque evolution \citep{Shi:2011us,Noble:2021vfg,Avara:2023ztw}. 
Qualitatively, the inner cavity was either excised or remained in a state resembling broadly previous hydrodynamical works.

On the other hand, magnetic fields in single accreting black holes have
been shown to cause drastic changes to the accretion flows \citep{Yuan:2014gma,Davis:2020wea}.
In particular, inferences from observations with the Event Horizon
Telescope indicate, that black holes such as M87* and Sgr A* \citep{EventHorizonTelescope:2019pgp,EventHorizonTelescope:2021srq,EventHorizonTelescope:2022urf,Collaboration:2024myo} might
be in a magnetically arrested (MAD) state \citep{Igumenshchev:2003rt,Narayan2003,Tchekhovskoy:2011zx}. In contrast to
advection dominated standard and normal accretion (SANE) flows, magnetically arrested flows are halted
by a strong magnetosphere around the black hole, essentially establishing
a cavity stabilized by magnetic pressure, which the flow can only cross via interchange-like instabilities
pushing the field locally into the accretion disk \citep{Tchekhovskoy:2011zx,Porth:2020txf,Ripperda:2021zpn}. 
Accretion then proceeds in episodic bursts, with the initial development of Rayleigh-Taylor fingers \citep{Zhdankin:2023wch} ( see
also, e.g., \citealt{Kulkarni:2008vk,Takasao:2022glf} for similar dynamics in T Tauri stars), before full magnetic flux tubes are ejected into the disk,
and cause an asymmetric burst of accretion on the black hole \citep{Porth:2020txf,Ripperda:2021zpn,Chatterjee2022}. This inflow of magnetic flux re-establishes a magnetically dominated zone, pushing out
the disk and resetting the magnetically arrested accretion cycle \citep{Tchekhovskoy:2011zx,Ripperda:2021zpn}.
Most importantly, angular momentum transport onto the black hole then mainly happens during the
flux eruptions, and itself is episodic, magnetically driven, and in parts outward-directed
\citep{Chatterjee2022}, and can spin down the black hole significantly \citep{Lowell:2023kyu}. At the same time, it has been suggested that magnetic flux eruptions
into the disk are mainly loaded with non-thermal electron-positron pairs and
can potentially power X-ray transient emission \citep{Dexter:2020cuv,Scepi:2021xgs,Zhdankin:2023wch,Hakobyan:2022alv,Vos:2023day} with implications for X-ray flares observed in the galactic center \citep{Gravity2018,Gravity2021}. 
It is important to point out that the picture painted here has been
mainly investigated for accreting black holes, although it is generic and has also
been found for accreting young proto-stars \citep{Kulkarni:2008vk,Takasao:2022glf} and for low-mass neutron star X-ray
binaries \citep{Parfrey:2023swe,Murguia-Berthier:2023fji} (see also \citealt{1976ApJ...207..914A,1978ApJ...219..617S,Spruit:1995fr}).\\

Motivated by recent progress in supermassive black hole binary formation
(e.g., \citealt{Shi2024}) and findings that large-scale accretion flows on parsec scales may well be very strongly magnetically dominated \citep{Hopkins2023,Hopkins2024,Guo2024}, it seems possible that a strongly magnetized state inside the circumbinary disk cavity could form, with accretion properties similar to magnetically arrested accretion flows on black hole horizon scales.\\

In this work, using massively parallel GPU-enabled simulations, we
demonstrate that the picture of magnetically arrested accretion flows around black holes naturally carries over to circumbinary disks, albeit with modifications introduced by the strong tidal potential of the binary.
In particular, we show that the circumbinary disk cavity can be rapidly filled
with vertical magnetic flux, establishing a magnetically arrested accretion state.
We further identify Rayleigh-Taylor fingers, flares, and flux eruptions, which we
find to reshape the cavity size and transport most of the angular momentum between the binary inside
the cavity and the outer circumbinary disk.
We also provide a detailed analysis of the angular momentum transport between the disk and the binary, finding that the magnetically arrested state has the potential to lead to an overall shrinking of the orbit.

\section{Methods}
We study the dynamics of a circumbinary accretion disk
endowed with a strong magnetic field.  To this end, we model the binary as
Newtonian point masses on a fixed Keplerian orbit, and the dynamics of the gas
flow using ideal magnetohydrodynamics \citep{Stone:2008mh}. The fluid is described by a
mass density, $\rho$, velocity, $\boldsymbol{v}$, internal energy density
$e$, pressure $P$, and magnetic field, $\boldsymbol{B}$. 
The evolution of the system then follows \citep{Stone2020},
\begin{align}
 \frac{\partial \rho}{\partial t}+\nabla \cdot(\rho \boldsymbol{v}) &=s_{\rho}\,, \\
 \frac{\partial(\rho \boldsymbol{v})}{\partial t}+\nabla \cdot(\rho \boldsymbol{v} \boldsymbol{v} + \bar{P} \mathbb{I} - \boldsymbol{B}\boldsymbol{B})&=\boldsymbol{s_p}-\rho \nabla \Phi\,, \\
 \frac{\partial E}{\partial t}+\nabla \cdot[(E+\bar{P}) \boldsymbol{v} - \left(\boldsymbol{B}\cdot \boldsymbol{v}\right) \boldsymbol{B}]&=s_E-\rho \boldsymbol{v} \cdot \nabla \Phi\,,\\
 \frac{\partial \boldsymbol{B}}{\partial t}-\nabla \times [ \boldsymbol{v} \times \boldsymbol{B} ]&=0\,,
\end{align}
where we have defined the conserved energy, $E$, and effective pressure
$\bar{P}$,
\begin{align}
    E = e + \frac{1}{2} \rho v^2 + \frac{1}{2}B^2\,,\\
    \bar{P} = P + \frac{1}{2}B^2\,.
\end{align}

The gravitational potential of the binary is modeled as 
\begin{align}
  \Phi = 
  - \frac{GM_1}{(\left|\boldsymbol{r}-\boldsymbol{r}_1\right|^2 + \epsilon^2)^{1/2}}
  - \frac{GM_2}{(\left|\boldsymbol{r}-\boldsymbol{r}_2\right|^2 + \epsilon^2)^{1/2}}\,,
\end{align}
where $M_i$ and $\boldsymbol{r}_i$ are the mass and location of the binary
constituents respectively, $G$ is the gravitational constant, $\boldsymbol{r}$
is the radial coordinate to the center of mass of the binary, and $\epsilon$ is a small number. We further adopt a unit convention of $G M=a=1$ in our simulations, and focus in an equal mass system, $M_1=M_2$.

Since we model the binary dynamics on scales of the orbital separation,
$a$, which are much larger than the physical extent of the binary
constituents (e.g., the horizon scale in the case of supermassive black hole binaries), we require an effective boundary/sink prescription.
While the type of sink can impact the formation of mini-disks and their
dynamics (see, e.g., \citet{Dittmann:2021wzj} for a detailed discussion), we here mainly focus on the evolution of the cavity, and adopt a
simple prescription removing matter at the local Keplerian rate,
 $\boldsymbol{s_p} = - \Omega_{K,i}\, \rho \left(\boldsymbol{v} - \boldsymbol{v}_i\right)$, 
 ${s_\rho} = - \Omega_{K,i}\, \rho$,  
 within a sink radius
 of $r_{\rm sink}= 0.07 a$, where $\Omega_{K,i} =
 (GM_i)^{1/2}/(\left|\boldsymbol{r}-\boldsymbol{r}_i\right|^2 +\epsilon^2)^{3/2}$
is the Keplerian frequency of each sink, 
$\epsilon=0.05a$ is a small number acting as Plummer gravitational softening, and $\boldsymbol{v}_i$ the velocity of the sink
on a fixed Keplerian orbit.
The form of ${s_E}$ follows from the density and momentum sink using the evolution equations.
 We further model the system as being locally isothermal,
 with the temperature being fixed to $T= c_s^2 = \left|\Phi\right|/\mathcal{M}^2$,
 where $c_s^2$ is the speed of sound and $\mathcal{M}=10$ is the sonic Mach number (or having disk aspect ratio $h=1/\mathcal{M}=0.1$).
 The surface density profile is initialized following \citealt{Duffell2024,Munoz:2020azx}, with the vertical structure being given by assuming constant temperature along the vertical coordinate.
 We initialize the magnetic field as purely poloidal by adopting a vector potential
 $A_\phi = R^2 A_0 {\max} \left(\rho -
 0.04\,\rho_{\max}, 0\right)^2$, where $R$ is the cylindrical radius, $\rho_{\max}$ the maximum mass density inside the initial disk, and $A_0$ is chosen to ensure an initial disk
 magnetization parameter $\beta = 2 P /B^2 = 20$. We further adopt a numerical floor value of $\beta_{\rm floor} = 10^{-3}\,.$ 
 \\

 We numerically integrate the above system using the public GPU-enabled
 version, \texttt{AthenaK}\footnote{\url{https://github.com/IAS-Astrophysics/athenak}}, of the \texttt{Athena++} code \citep{Stone2020} based on the \texttt{Kokkos} library
 \citep{Trott2021}. 
 In this work, we solve the Newtonian MHD system using piecewise parabolic reconstruction \citep{Colella1984}, an HLLD Riemann solver \citep{Miyoshi2005}, and a
 constraint transport algorithm \citep{Gardiner2008} for the divergence-free magnetic field evolution.
 Timestepping is done using a second-order accurate Runge-Kutta stepper.

 To perform the simulation, we adopt a Cartesian grid with 6 levels of
 static mesh refinement. The levels decrease in extent by a factor two each from the outer (outflow) domain boundary inward, which is located at $\pm 80 a$. We adopt a finest resolution of $\Delta x \simeq 0.0065 a$. We then numerically integrate
 the system on 1,260 V100 GPUs on the DOE OLCF Summit system for 230 binary orbits. 
  Overall, the cost of the simulation presented here is about 135,000 V100 GPU-hours.

\begin{figure*}
    \centering
    \includegraphics[width=0.95\linewidth]{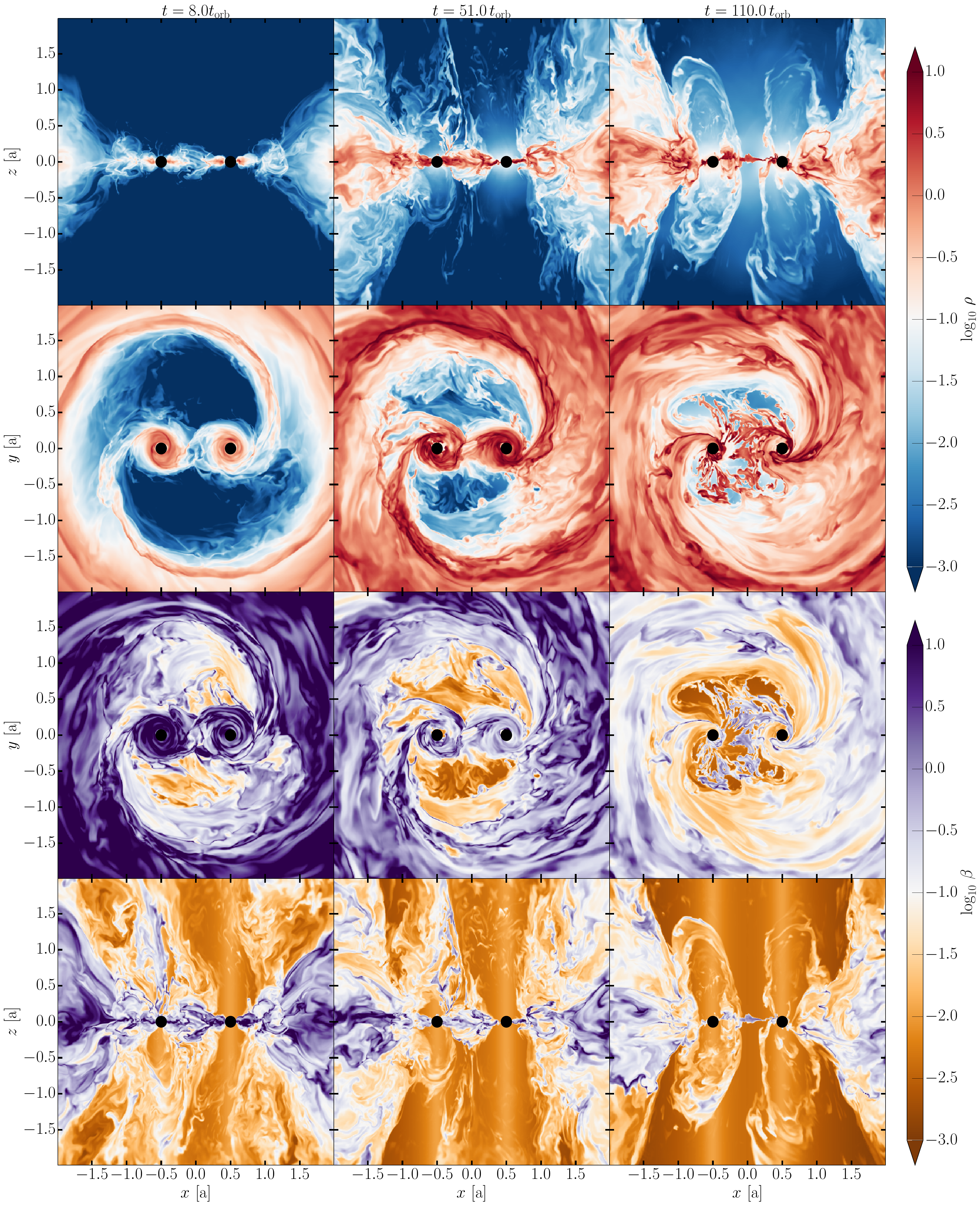}
    \caption{Initial circumbinary evolution towards a magnetically arrested state over the first 110 binary orbits.\\ {\it (Top and bottom)} Meridional slices along the orbital axis, {\it (center)} equatorial slices. Shown in color are the mass density, $\rho$, and magnetization parameter $\beta$. Times, $t$, are stated relative to the orbital period, $t_{\rm orbit}$, whereas distances are stated relative to the binary separation, $a$.} 
    \label{fig:initial}
\end{figure*}

\begin{figure*}
    \centering
    \includegraphics[width=0.95\linewidth]{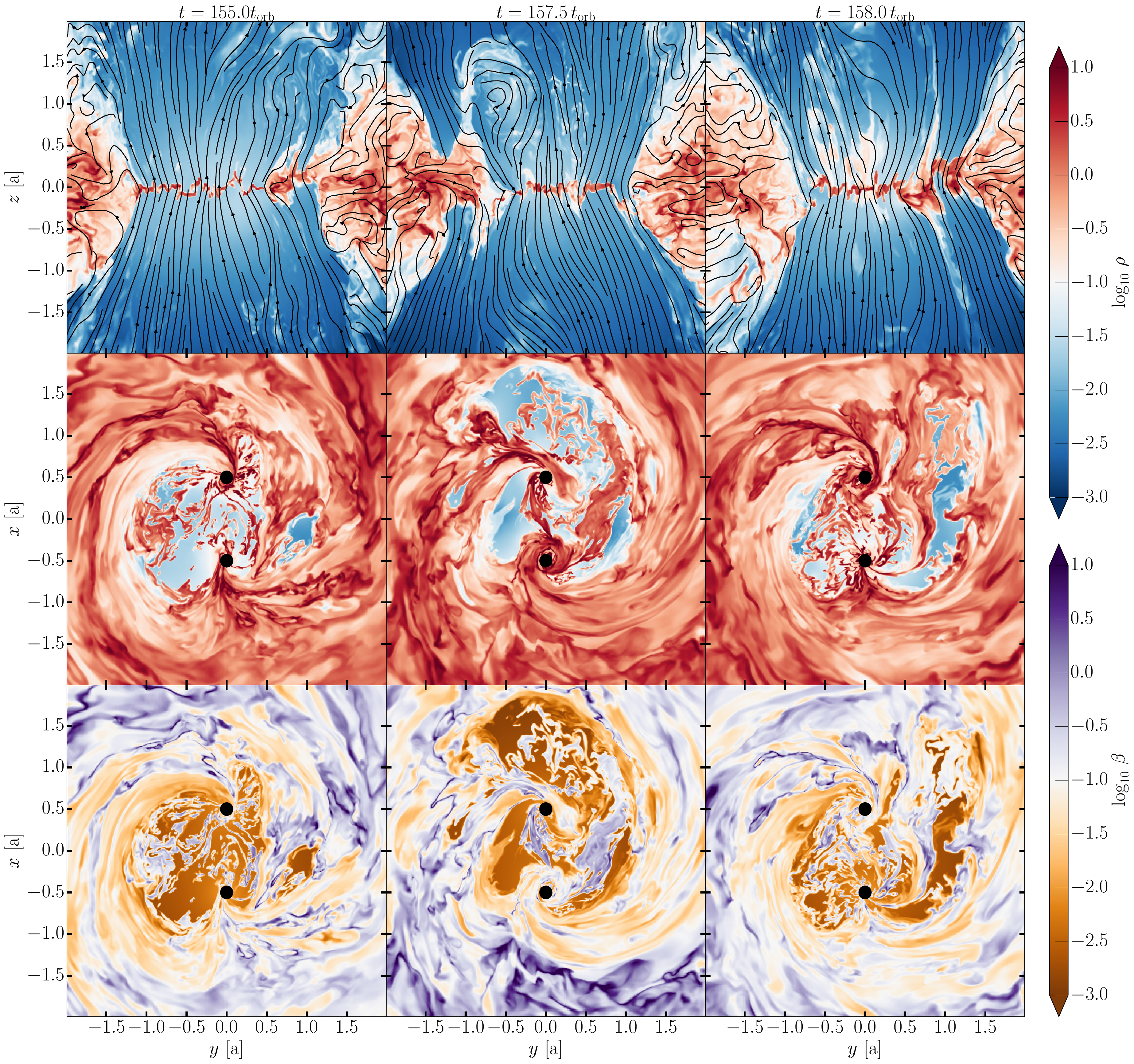}
    \caption{Circumbinary accretion cycle in the magnetically arrested state. {\it (Left)}  Quiescent state, strong vertical magnetic fields (black streamlines) push out the accretion flow and support it. The cavity truncates well inside the inner Lindblad radius. We can also see a previous eruption epsiode to the right of the central cavity.  {\it (Center)} Interchange instabilities trigger accretion streams through the cavity onto the mass sinks, with the cavity being highly distorted and expanded. We can also see that a reconnection-driven magnetic flare is launched vertically outwards. {\it (Right)} This leads to an ejection of vertical net flux from the cavity in the circumbinary disk. The ejected 
    (magnetic pressure dominated) 
    flux tube gets strongly sheared. 
    Shown in color are the mass density, $\rho$, and magnetization parameter, $\beta$, with the top and bottom rows showing the meridional and equatorial planes, respectively. Times, $t$, are stated relative to the orbital time period, $t_{\rm orb}$.}
    \label{fig:mad_cycle}
\end{figure*}

\section{Results}
We present our results on magnetically arrested circumbinary disk accretion in the following way.
First, we discuss the initial build-up of magnetization inside the cavity and its transition to a magnetically arrested state. We then describe the accretion cycle by means of magnetic flux eruptions, before concluding with an in-depth discussion on angular momentum transport.\\

\subsection{Evolution towards a magnetically arrested state}
We begin by detailing the overall evolution of the structures inside the cavity, which is shown in Fig. \ref{fig:initial}. 
Initially, the cavity is empty and begins to fill rapidly through tidal accretion streams, as can be seen after eight orbital periods. This phase is not unlike the early onset of two-dimensional circumbinary accretion (e.g., \citealt{Munoz2019,Duffell2024}), or recent three-dimensional simulations with an outer accretion flow \citep{Moody2019,Avara:2023ztw,Pieren2023,2024ApJ...964..133M}. In this stage, rapid mini-disk formation sets in, with the outer disk subsequently becoming unstable to the magnetorotational instability (MRI) \citep{Balbus:1991ay} and turbulent. The magnetization, $\beta = 2 P / B^2 \gg 1$, in most parts of the cavity including inside the mini-disks. Overall, despite the more complex effective spatially varying viscosity (see \citet{2024ApJ...964..133M} for a detailed discussion), this stage of our evolution can be characterized as mainly hydrodynamical.\\
As the evolution proceeds, accretion from the circumbinary disk advects substantial amounts of magnetic flux into the cavity, leading to a build-up of net magnetic flux. We can see that already after 51 orbits, some parts of the cavity are entirely magnetically dominated, with $\beta \ll 1$. While the tidal accretion streams are still predominantly present, the mini-disks start to exhibit turbulent features, with their magnetization level approaching equipartition values, $\beta \simeq 1$. The outer disk has become strongly turbulent at the inner cavity wall, with strong magnetically driven outflows present above the sink cells. These jet-like features are consistent with strong net vertical flux inside the cavity.
In this transitional state, the flow is still largely hydrodynamical.\\
\begin{figure}[t]
    \centering
\includegraphics[width=\linewidth]{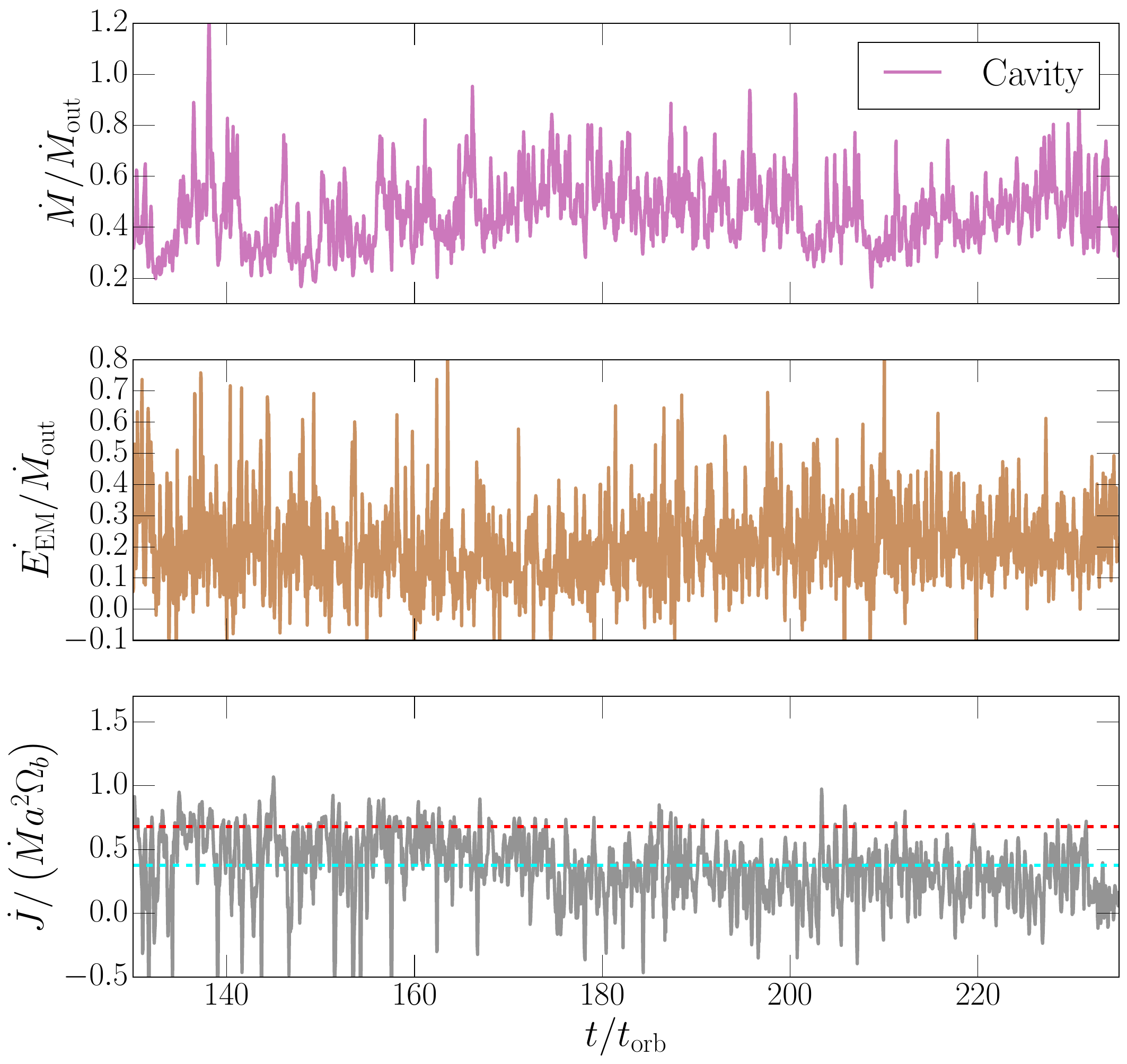}
    \caption{Mass, $\dot{M}$, electromagnetic energy, $\dot{E}_{\rm EM}$, and angular momentum, $\dot{J}$, fluxes into the cavity, normalized to the mass accretion rate, $\dot{M}_{\rm out}$, at large distances. The time series is shown relative to the orbital period, $t_{\rm orbit} \equiv 2\pi/\Omega_b$. The red dashed line denotes the expected value of $\left<\dot{J}/\dot{M}\right> \simeq 0.68\, a^2 \Omega_b$ for a converged non-eccentric, equal mass, vertically integrated hydrodynamic accretion flow \citep{Munoz2019}, the cyan dashed line the threshold value,$\left<\dot{J}/\dot{M}\right> \simeq 3/8\, a^2 \Omega_b$, for the binary orbit to shrink \citep{Miranda2017,Lai2023}.}
    \label{fig:mdot}
\end{figure}

After about 100 orbits, however, the flow structure inside the cavity has drastically changed. The continued inflow of magnetic flux into the cavity has lead to a regime that is largely magnetically dominated, with $\beta \simeq 10^{-3}$ almost everywhere inside the cavity. Rather than having a turbulent accretion flow at the inner boundary wall, we can see a very clear and sharp delineation between the cavity and the accretion flow in the meridional plane. This state is reminiscent of a magnetically arrested accretion flow onto a single black hole \citep{Tchekhovskoy:2011zx,Begelman:2021ufo,Chatterjee2022}. There, the magnetic field pushes out the disk, with mass accretion in the quiescent state being confined to a thin current sheet in the equatorial plane of the black holes \citep{Ripperda:2021zpn}. Different from hydrodynamical simulations in which the radial location of the cavity is fully determined by the viscous and gravitational torque balance, here the joint torque from the magnetic field and gravitational interaction balances the torque from MRI turbulence in the circumbinary disk, causing a smaller cavity size.
We observe the presence of a similar equatorial current sheet in our simulations (Fig. \ref{fig:mad_cycle}, top row). 

\subsection{Magnetic flux eruption cycle }
\begin{figure}[t]
    \centering
\includegraphics[width=\linewidth]{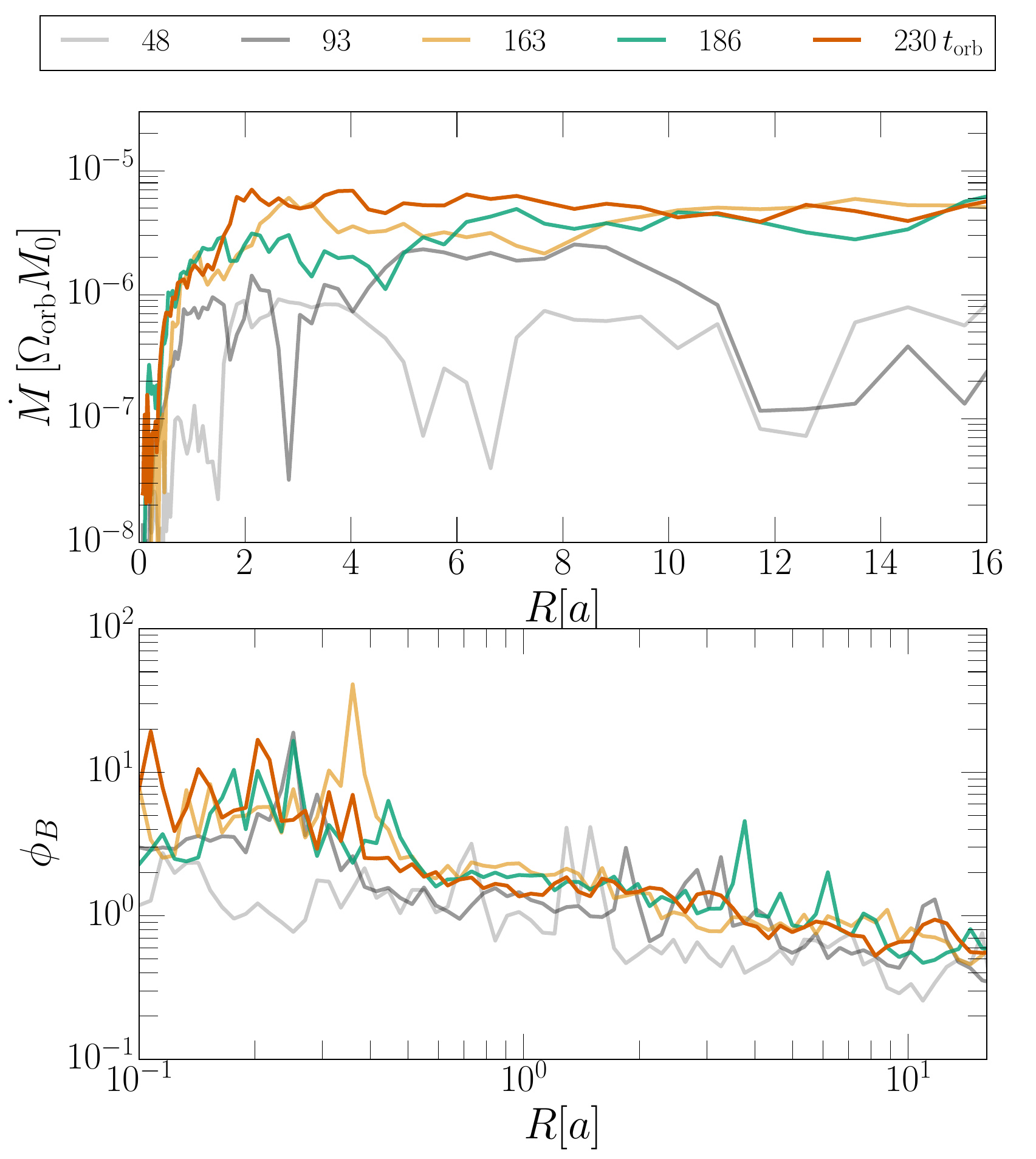}
    \caption{Radial profile of the mass accretion rate, $\dot{M}$, and the normalized magnetic flux, $\phi_B(r)=\Phi_B (r)/\sqrt{\dot{M} r^2}$, where $\Phi_B(r)$ is the magnetic flux at cylindrical radius $R$, given relative to the binary separation, $a$.
    $\Omega_{\rm orb}$ and $M_0$ are the binary orbital frequency and initial disk mass, respectively.
    Different times are shown with different colors, as denoted in the legend.}
    \label{fig:mdot_r}
\end{figure}
One characteristic feature of magnetically arrested accretion flows around single black holes is an intermittent phenomenon regarding the exchange of mass and angular momentum \citep{Tchekhovskoy:2011zx,Ripperda:2021zpn,Chatterjee2022}.
The system features two accretion states: a quiescent state and an eruption state. We show this cycle for our circumbinary disk simulation in Fig. \ref{fig:mad_cycle}.

In the quiescent state, strong magnetic fields prevent the cavity from accreting mass except through tidal accretion streams. 
However, our simulations indicate that the mass and angular momentum carried by these accretion streams might be insufficient to provide for permanent mini-disk formation.
We caution that the size of the mini-disks (if any during this accretion phase) that we find in our simulation is subject to resolution effects and magnetic flux accumulation inside the sink, which likely will not influence the cavity wall dynamics we investigate here. %

At the end of the quiescent state, mass accumulates at the inner cavity wall on a viscous timescale of the outer accretion flow. 
Eventually the growth of an interchange instability \citep{Spruit:1995fr} leads to the inward accretion of Rayleigh-Taylor fingers into the cavity, which does not coincide with the tidal accretion streams.
This appearance of Rayleigh-Taylor fingers is not unlike accretion states that have been found for T Tauri stars \citep{Kulkarni:2008vk,Takasao:2022glf}, low mass X-ray binaries \citep{Parfrey:2023swe,Murguia-Berthier:2023fji}, and magnetically arrested black holes \citep{Begelman:2021ufo,Zhdankin:2023wch}. 
These magnetic flux eruptions lead to outward-moving vertical magnetic flux tubes that get ejected into the accretion disk (Fig. \ref{fig:mad_cycle}, left and right column). In analogy with the accretion cycle for magnetically arrested black hole flows \citep{Chatterjee2022,Begelman:2021ufo}, these low-density, high magnetic pressure flux eruption regions are asymmetric and will lead to net torques on the system, as we will detail in Sec. \ref{sec:torques}. 
In this state, mass accretion is enhanced, leading to a transient formation of a mini-disk as magnetic pressure support on the outer cavity wall is released.
Concurrently, the rapid accretion of magnetic flux leads to reconnection events that can trigger the ejection of large plasmoids/flares (see Fig. \ref{fig:mad_cycle}, top center panel). These have also been found in magnetically arrested   black hole disks, and have been suggested as an explanation for X-ray flares in the galactic center \citep{Porth:2020txf,Ripperda:2020bpz,Nathanail:2021jbn}.

\begin{figure}
    \centering
\includegraphics[width=\linewidth]{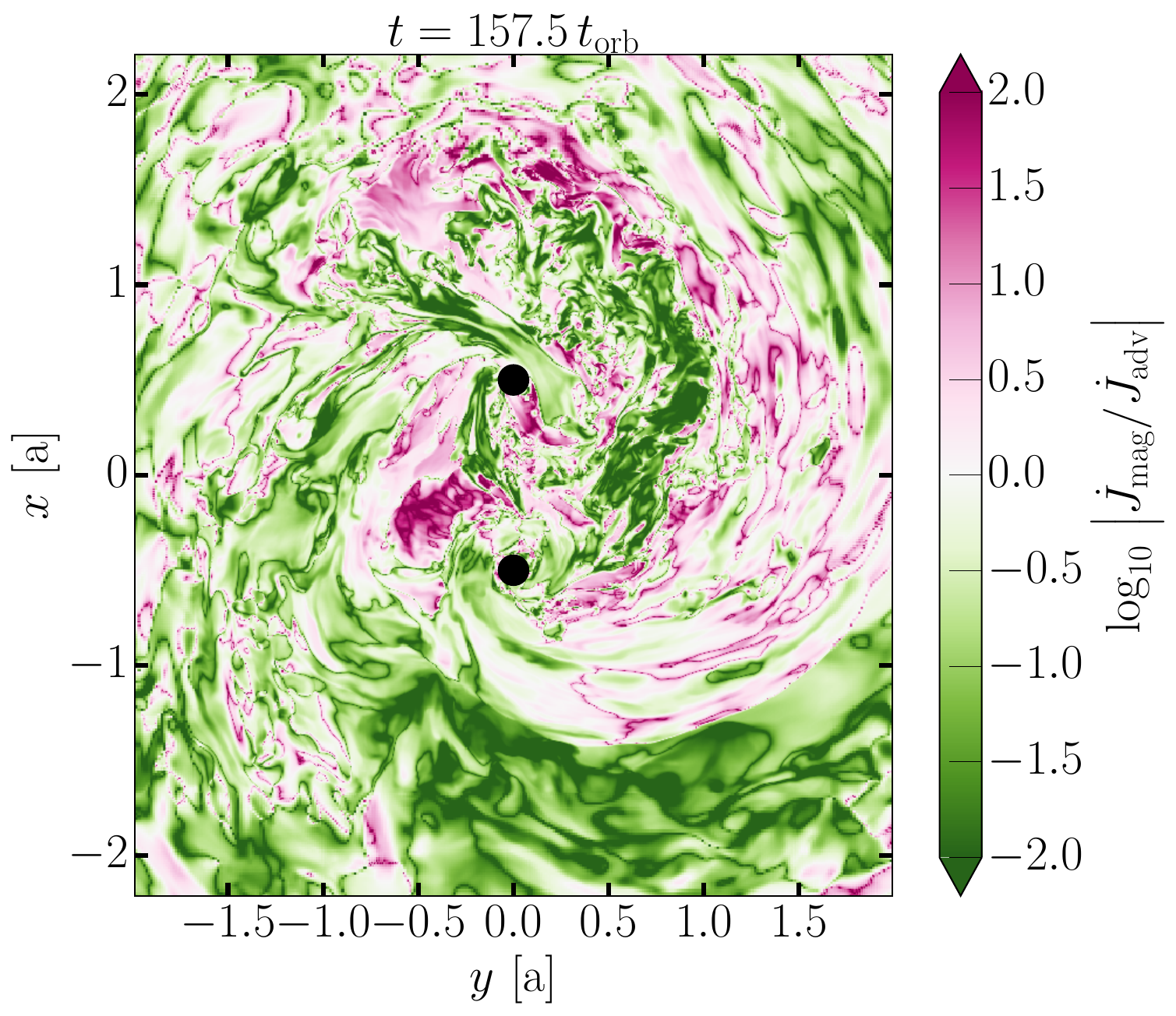}\\
\includegraphics[width=\linewidth]{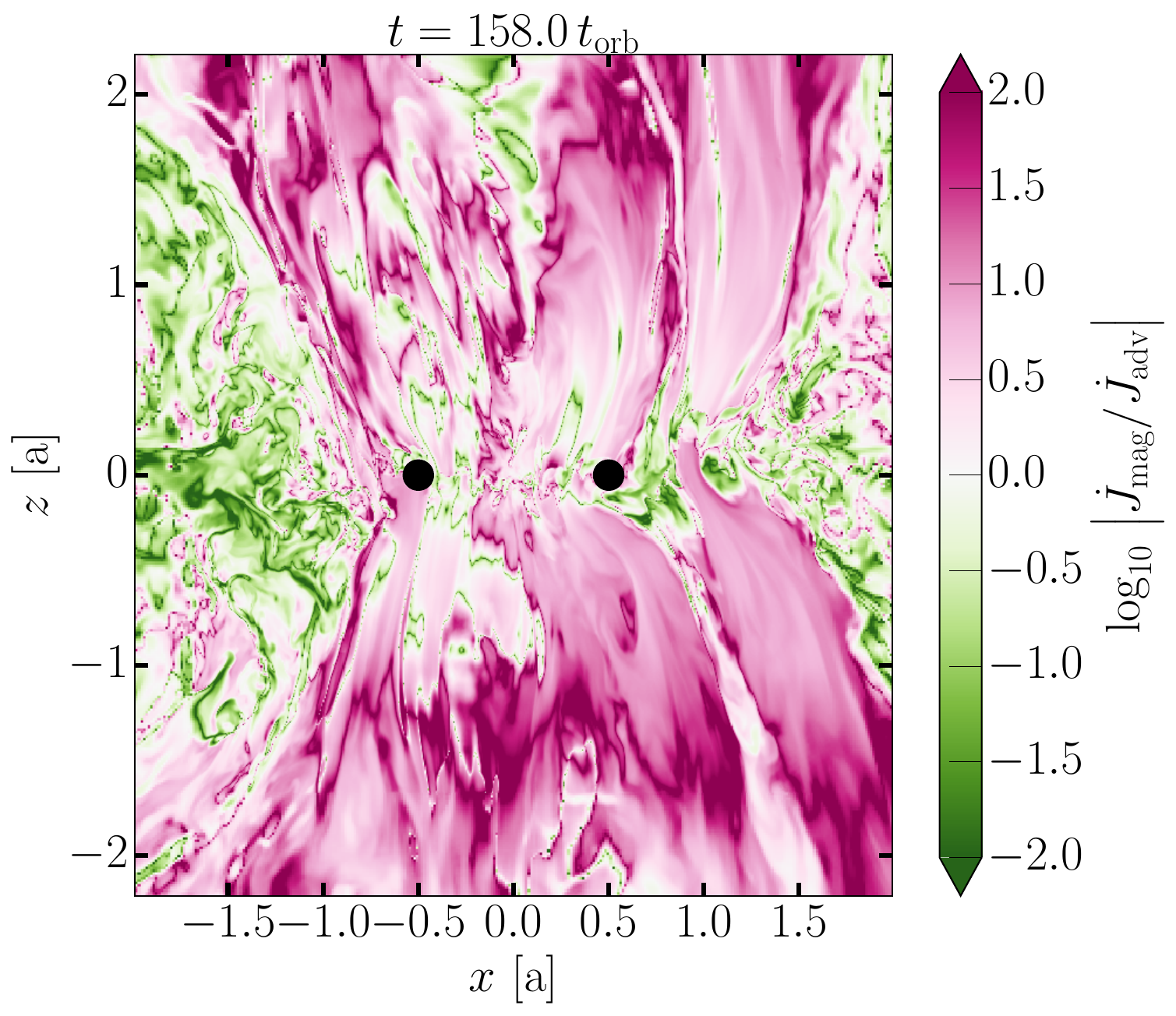}\\

    \caption{Angular momentum transport in the circumbinary disk during a flux eruption.
    Shown is the ratio of the magnetic, $\dot{J}_{\rm mag}$, to advected, $\dot{J}_{\rm adv}$, angular momentum fluxes.
    {\it (Top)} Flux eruption in the equatorial plane. Angular momentum is largely transported outwards due to magnetic stresses (pink regions). {\it (Bottom)} Angular momentum fluxes inside a flux tube (right). As the flux tube gets ejected into the disk ($ 1<x/a<1.5$), it drives a wind carrying angular momentum outwards.}
    \label{fig:torque}
\end{figure}

\begin{figure*}
    \centering
    \includegraphics[width=0.45\linewidth]{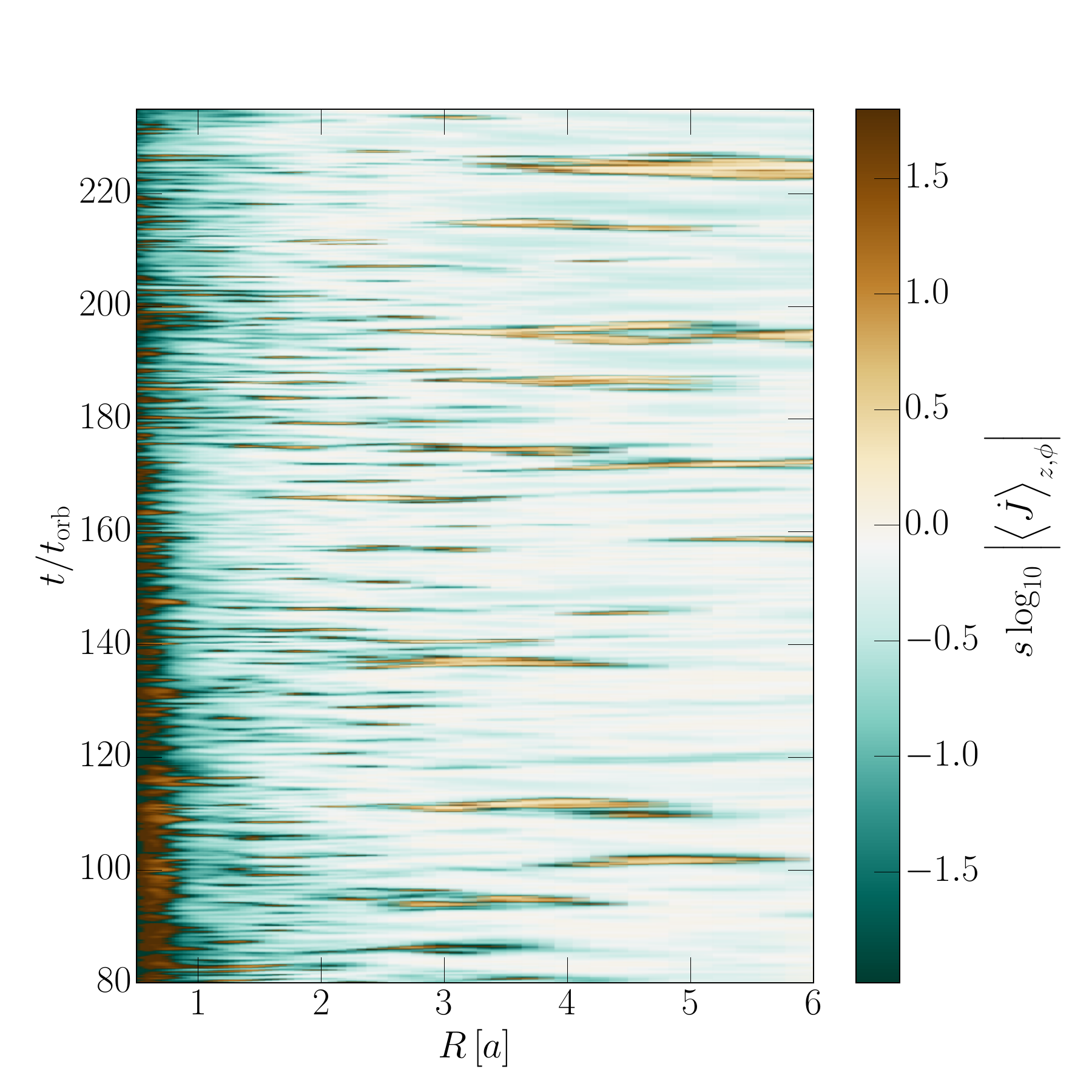}
\includegraphics[width=0.45\linewidth]{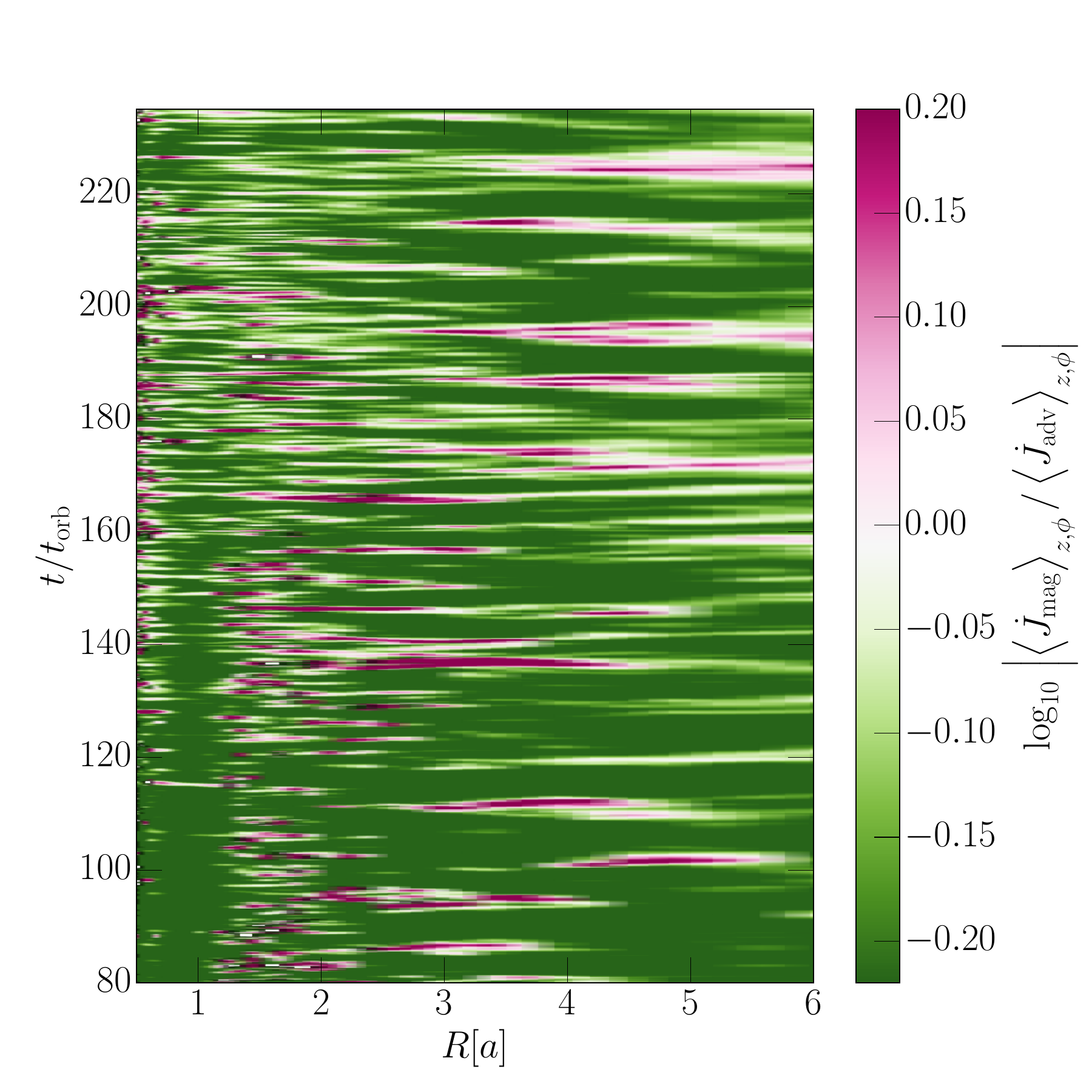}
    \caption{Azimuthally and vertically integrated radial angular momentum fluxes as a spacetime diagram. {\it(Left)} Total radial angular momentum flux, $\left<\dot{J}\right>_{z,\phi}$. $s=+1\, (\rm outwards) / -1\, (\rm inwards)$ indicates the sign of the flux. {\it(Right)} Ratio of the magnetic, $\left<\dot{J}_{\rm mag}\right>_{z,\phi}$, to the advective, $\left<\dot{J}_{\rm adv}\right>_{z,\phi}$, contribution of the radial angular momentum flux. The cylindrical radius, $R$, is shown in units of the binary separation, $a$. Times, $t$, are shown relative to the orbital period, $t_{\rm orbit}$. }
    \label{fig:jdot}
\end{figure*}

After the magnetic flux tubes got ejected into the disk and sufficient magnetic flux has been re-accreted into the cavity, the cavity restores its magnetically dominated state, resetting the cycle and suppressing Rayleigh-Taylor instabilities, at least intermittently. Meanwhile, the ejected flux tubes get sheared and mixed into the disk (Fig. \ref{fig:mad_cycle}, right column).

We can also quantitatively study the dynamics of the cavity. To this end, we compute the mass accretion rate, $\dot{M}$, into the cavity, which we approximate by a sphere around the binary's center of mass with radius $r\simeq 0.6 a$. 
The cavity we observe is generally smaller than in hydrodynamic simulations \citep{Munoz2019}, although we caution that our accretion state and cavity size may not yet be fully converged.
As shown in Fig. \ref{fig:mdot}, the magnetically arrested flow leads to periodic mass accretion patterns, which vary by a factor of 3-4, when normalized to the mass accretion rate, $\dot{M}_{\rm out}$, at larger distances ($r\simeq 3.5 a$).
Indeed, we can spot clear accretion cycles matching the flux eruptions with a period of $3-5$ orbits as seen in Fig. \ref{fig:mad_cycle}. At the same time, we find that the mass accretion rate is already well converged out to radii of $r \simeq 15\, a$ (Fig. \ref{fig:mdot_r}). This likely implies that the flow we observe is roughly in inflow equilibrium.\\
In the black hole accretion community, commonly the dimensionless flux $\phi_B (r) = \Phi_B(r) / \sqrt{|\dot{M}| r^2}$, where $\Phi_B(r)$ is the magnetic flux computed over a a sphere of radius $r$, commonly the horizon \citep{Tchekhovskoy:2011zx}. In a magnetically arrested state, a large value of $\phi_B \gg 1$ implies that energy in excess of the accreted mass energy can be extracted from the black hole \citep{Blandford:1977ds,Tchekhovskoy:2011zx}.  We therefore compare $\phi_B$ as a function of radius, $r$, for different times (Fig. \ref{fig:mdot_r}). We can see that $\phi_B \gg 10$ locally inside the cavity. At the same time the outer disk region is at $\phi_B \lesssim 1$.
We also compare the net flux of electromagnetic energy $\dot{E}_{\rm EM}$ into the cavity (Fig. \ref{fig:mdot}). Correlating this with the mass accretion rate, we find that spikes in the electromagnetic energy flux usually precede eruption events, consistent with the necessary requirement of replenishing the magnetic flux inside the cavity after a successful eruption.

\subsection{Angular momentum transport via flux eruptions}\label{sec:torques}

One important aspect of circumbinary disk accretion is the question of angular momentum transport, or conversely, whether interactions with the gaseous disk make the binary orbit spin up or shrink for equal mass systems \citep{Munoz2019,Duffell:2019uuk,Duffell:2024fwy}.

Magnetically arrested accretion flows can feature yet another angular momentum transport channel. As has been shown by \citet{Chatterjee2022}, flux eruptions themselves are able to transport angular momentum outwards.
This process is intrinsically intermittent and quasi-periodic, mainly influenced by the magnetic accretion cycle.
As we will show in the following, this picture seems to largely carry over to magnetically arrested accretion flows.
Conversely, the periodicity of angular momentum transport presented in hydrodynamical simulations is governed by tidal accretion streams and a slowly precessing eccentric cavity.

We quantify the amount of angular momentum transport by considering the transport equation for angular momentum in cylindrical coordinates, which can be obtained from the Euler equations (see the discussion of vertically integrated angular momentum transfer in \citet{Miranda2017}, and the three-dimensional version in \citet{2022ApJ...925..161S}). The conserved angular momentum density, $j = R^2 \rho v_\phi$, is then given by,
\begin{align}
    &\partial_t j + \partial_R \left( j v_R - R^2 B_R B_\phi \right) + \partial_\phi \left( {j} - R B_\phi B_\phi + R \bar{P}\right)\nonumber\\ 
    &+ \partial_z \left(j v_z  - R^2 B_z B_\phi\right) + R \rho\, \partial_\phi \Phi = 0\,,
\end{align}
where $R$ as the cylindrical radius, $z$ as the height coordinate, and $\phi$ the azimuthal coordinate. 
 
We compute the effective angular momentum fluxes as integrals on cylindrical surfaces located at the binary's center of mass, i.e.,
\begin{align}
    \partial_t \left<j\right>_{z,\phi} + \partial_R \left( \left<\dot{J}_{\rm adv}+ \dot{J}_{\rm mag}+ \dot{J}_{\rm grav}\right>_{z,\phi} \right) = 0\,,
\end{align}
where we have defined the $\left<\ldots\right>_{z,\phi}$ as the $z-$ and $\phi-$integral over the cylinder, and 
\begin{align}
    \dot{J}_{\rm adv} &= R^2  \rho v_\phi v_R\,,\\
    \dot{J}_{\rm mag} &= -R^2  B_R B_\phi\,,\\
    \dot{J}_{\rm grav} &=  \int_0^R  d r'  r' \rho\, \partial_\phi \Phi \,,
\end{align}
as the advective and magnetic angular momentum flux, respectively. We further define the total averaged angular momentum flux as $\dot{J} = \dot{J}_{\rm adv} + \dot{J}_{\rm mag} + \dot{J}_{\rm grav}$. 
Following \citet{Chatterjee2022}, we define the sign of the angular momentum flux such that $\dot{J} > 0$ corresponds to an outward flux of angular momentum, whereas $\dot{J} < 0$ corresponds to an inward one.

Using this decomposition, we can analyze the torque carried by a flux eruption event (Fig. \ref{fig:torque}). Focusing on the same accretion cycle shown in Fig. \ref{fig:mad_cycle}, we can see that the torque inside the flux eruption is entirely magnetically dominated. In magnetically dominated regions of the erupted flux tube, magnetic angular momentum fluxes are up to 100-times larger than the advected hydrodynamical fluxes. Outside of the eruption region, and also inside the Rayleigh-Taylor fingers, the angular momentum fluxes are largely hydrodynamical. 
The ejected flux tube carries angular momentum into the disk, which will eventually drive a wind/outflow due to the strong vertical net flux. This behavior is consistent with flux eruptions around single black holes \citep{Chatterjee2022,Lowell:2023kyu}.

We can further quantify the flux transport behavior throughout the evolution of the system over several eruption cycles. We show the resulting temporal evolution of the angular momentum flux in Fig. \ref{fig:jdot}, focusing on the evolution after the first 100 orbits when the system has become magnetically arrested. We can see that consistent with the appearance of magnetically arrested accretion cycles (over several binary orbits) the system features periodic outbursts of radial angular momentum flux. This flux starts roughly at or slightly outside the stationary cavity radius, $r\simeq a$, and extends outward to several binary distances. These fluxes are also outward-pointing. Indeed, when comparing the strength of the magnetically driven to the advected angular momentum flux (Fig. \ref{fig:jdot}, right panel), we find that these outbursts are entirely magnetically driven, consistent with the flux eruption picture (Fig. \ref{fig:torque}), and with a similar analysis in magnetically arrested black hole disks \citep{Chatterjee2022}.

It can be shown that for an equal mass non-eccentric accreting binary, such as the one consider here, the evolution of the orbital separation, $a$, should follow \citep{Miranda2017,Lai2023},
\begin{align}
   \frac{\dot{a}}{a} = 8 \left(\ell_0 - \frac{3}{8}\right) \frac{\langle\dot{M}\rangle}{M}\,, 
\end{align}
where $\ell_0 \equiv \langle \dot{J}\rangle/\left(\langle\dot{M}\rangle a^2 \Omega_b\right)$, and $\Omega_b$ is the Keplerian frequency of the binary. From this equation, we can see that for $\ell_0 < 3/8$, the binary separation would shrink. Hydrodynamical simulations in two-dimensions imply $\ell_0 \simeq 0.68$ \citep{Munoz2019}, whereas three-dimensional simulations find $\ell_0 = 0.55$ \citep{Moody2019}.
Both these values lead to an eventual growth of the binary orbit.

On the other hand, we find that during flux eruptions $\ell_0$ can instantaneously drop below the threshold value, $\ell_0 < 3/8$, or even become negative (Fig. \ref{fig:mdot}, bottom panel). Correlating these patterns with those of the mass accretion rate, $\dot{M}$, we find that this behavior correlates directly with the magnetically arrested accretion cycle. In the quiescent phase, accretion of angular momentum onto the binary appears to be hydrodynamical and approaches approximately the equilibrium value, $\ell_0\simeq 0.68$, found in two-dimensional studies.
During flux eruptions when transport is entirely magnetically driven, angular momentum can move outwards.\\
We caution, that while our findings qualitatively match results for single black hole accretion flows \citep{Chatterjee2022,Lowell:2023kyu}, our overall flow is very likely not yet converged, and the cavity could still develop strong eccentricities \citep{Munoz2019}. A determination of which effect ultimately dominates total angular momentum transfer and budget in this system, will likely require substantially longer simulations than the ones presented here.

\section{Conclusions}

In this work, we have numerically investigated a magnetically arrested accretion state of a circumbinary disk.

Our simulations indicate, that just as for single accreting black holes, circumbinary accretion flows can become magnetically arrested. 
In this state, the inner cavity is filled with a strong magnetic flux regulating mass and angular momentum transport onto the binary \citep{Chatterjee2022}.
Magnetically arrested accretion states have been well studied in the black accretion literatures \citep{Narayan2003,Igumenshchev:2003rt,Igumenshchev:2007bh,Tchekhovskoy:2011zx}.
Different from these works, we find that the presence of tidal forces from the binary induces constantly present tidal streams, 
which, however, are periodically disrupted and reformed during the magnetically arrested accretion cycle.\\

Periodic magnetic flux eruptions through magnetic interchange instabilities \citep{Spruit:1995fr,Kulkarni:2008vk}, will then enhance mass transport inwards, and transiently drive angular momentum outwards. 
At the same time, this process triggers reconnection and the ejection of plasmoids and flares, akin to models for X-ray flares in the galactic center \citep{Porth:2020txf,Scepi:2021xgs,Ripperda:2020bpz,Ripperda:2021zpn}. The ejection of flux tubes into the disk is likely also accompanied with potential non-thermal emission \citep{Zhdankin:2023wch,Ripperda:2021zpn}.
The mixing of the ejected flux tubes with strong vertical net flux with the disk may also affect wind launching from the system \citep{Blandford:1982xxl,2018ApJ...857...34Z,2024MNRAS.528.2883Z}.
Furthermore, the strong magnetic field present in disk alters the cavity truncation radius, making it smaller than what is commonly observed in hydrodynamical flows \citep{Munoz2019}. This may have implications during the decoupling phase of the binary, potentially avoiding a drop of the mass accretion rate \citep{Noble:2021vfg,Avara:2023ztw}.

Most importantly for our context, magnetically arrested accretion flows around black holes have been shown to transport angular momentum outwards in magnetic flux eruption episodes \citep{Chatterjee2022}, although for thick magnetically arrested disks the main spin-down mechanism may originate from the jet \citep{Lowell:2023kyu}. We similarly find that magnetic flux eruption episodes carry angular momentum outwards, which can satisfy the criterion for the binary orbit to shrink. 
In this sense, magnetically arrested flows may offer a way to provide the necessary angular momentum loss for the binary's orbit to shrink. We, indeed, find prolonged binary cycles over which the orbit would be shrinking (see Fig. \ref{fig:mdot}).  However, we caution that our results are very likely not yet in a fully converged regime, with longer integration times being necessary to render a verdict on the actual net angular momentum transport rate.

While we find the period of the magnetically arrested circumbinary accretion cycle to be between $3-5$ orbits, the precise duration will likely depend on the amount of magnetic flux inside the accretion flow. Here we model the initial accretion disk as having largely poloidal flux, enabling the disk to easily replenish the amount of magnetic flux inside the cavity. 
Recent works on large-scale feeding onto supermassive black holes
indicate that the magnetic field at large distances may be toroidal, with the disk being largely magnetically and radiation dominated (e.g., \citealt{Hopkins2023,Hopkins2024}; see also \citealt{Cho:2023wqr,Guo2024}).
When accreting such magnetic fluxes, further dynamo action either inside the circumbinary or mini-disks may likely be necessary to produce the required net vertical fluxes for the cavity to make the flux eruption happen (e.g., \citealt{Liska:2018btr,Jacquemin-Ide:2023qrj,Sadanari2024}).
This will potentially prolong the timescale between magnetically arrested cycles, likely making this feature intermittent.

The cavity structure and mini-disk formation could also be affected by changes in the magnetic field topology or large-scale fluxes.
Additionally, we point out that our sink size is only $7\%$ of the binary separation, and does not include an effective resistivity to regulate magnetic flux accumulation in the sink. Further investigations will be required as to how this will affect the magnetic flux filling on the cavity.
Thermal effects in such disks may also naturally alter the circumbinary accretion state \citep{Wang2023}.
We plan to study the role of large scale magnetic fluxes in a future work.

Besides uncertainties in the (effectively initial) magnetic flux distribution, the development of an eccentric cavity \citep{Munoz2019} may likely alter the onset of the interchange instability and the periodicity of the flux eruption.
Quantifying this effect will require simulations such as the one presented here, over likely 10-times longer timescales, which is beyond the scope of this initial work, but well within in reach of modern GPU-based simulation approaches. 

\section*{Acknowledgments}
The authors are grateful for insightful discussions with Xue-Ning Bai, Manuela Campanelli, Luciano Combi, Alexander Dittmann, Philip F. Hopkins, Julian H. Krolik, Douglas N. C. Lin, Scott C. Noble, E. Sterl Phinney, Bart Ripperda, James M. Stone, Alexander Tchekhovskoy, and Zhaohuan Zhu. 
The simulations were performed on DOE OLCF Summit under allocation AST198. 
Additional simulations were done on DOE NERSC supercomputer Perlmutter under grant m4575.
HYW thanks for the support from B. Thomas Soifer Chair's Graduate Fellowship.
Part of this project was completed during the second LISA Sprint meeting at Caltech. The meeting was supported by the Jet Propulsion Laboratory Astronomy and Physics Directorate.
This research was supported in part by grant NSF PHY-2309135 to the Kavli Institute for Theoretical Physics (KITP).
This work was performed in part at Aspen Center for Physics, which is supported by National Science Foundation grant PHY-2210452.

\software{AthenaK \citep{Stone2020},
      Kokkos \citep{Trott2021},
	  matplotlib \citep{Hunter:2007},
	  numpy \citep{harris2020array},
	  scipy \citep{2020SciPy-NMeth}
}

\bibliography{inspire,non_inspire,cbd}

\end{CJK*}
\end{document}